\documentclass[conference, 9pt]{IEEEtran}
\IEEEoverridecommandlockouts
\usepackage{cite}
\usepackage{amsmath,amssymb,amsfonts}
\usepackage{algorithmic}
\usepackage{graphicx}
\usepackage{textcomp}
\usepackage{xcolor}
\def\BibTeX{{\rm B\kern-.05em{\sc i\kern-.025em b}\kern-.08em
    T\kern-.1667em\lower.7ex\hbox{E}\kern-.125emX}}
\begin{document}

\title{Dynamic DNNs Meet Runtime Resource Management on Mobile and Embedded Platforms\\

\thanks{These works were supported in part by the Engineering and Physical Sciences Research Council (EPSRC) under Grant EP/S030069/1. No data except those explicitly stated in the paper were generated during the study.}
}

\author{\IEEEauthorblockN{Lei Xun, Jonathon Hare, Geoff V. Merrett }
\IEEEauthorblockA{\textit{University of Southampton, UK}\\
\{lx2u16, jsh2, gvm\}@ecs.soton.ac.uk}
\and
\IEEEauthorblockN{Bashir M. Al-Hashimi}
\IEEEauthorblockA{\textit{King’s College London, UK}\\
bashir.al-hashimi@kcl.ac.uk}
}

\maketitle



Deep neural network (DNN) inference is increasingly being executed on mobile and embedded platforms due to low latency and better privacy. However, efficient deployment on these platforms is challenging due to the intensive computation and memory access.

Static model pruning is an effective method to reduce the model parameters (i.e. computation demand) at the cost of accuracy loss. Platform-aware model pruning \cite{yang2018netadapt} gradually reduces the number of parameters while measuring the latency, and stops pruning once the target latency is achieved in order to keep the accuracy as high as possible. However, the latency measurement is based on a fixed hardware configuration (e.g. CPU/GPU at maximum clock frequency), which is often unavailable at runtime due to thermal throttling or other concurrently executing applications sharing the hardware. The target latency could also change during different execution phases of the same application \cite{xun2020optimising}. Although a conservative hardware configuration can guarantee hardware availability at runtime, more parameters need to be pruned to achieve the same latency target, leading to a fixed model with lower accuracy not fully utilising available hardware.

Unlike static model pruning, which can only fit into a fixed latency target and hardware configuration, dynamic DNNs are trained to have sub-networks with different latency-accuracy trade-offs through either channel scaling \cite{yu2018slimmable, xun2019incremental} or layer scaling \cite{laskaridis2020hapi}. For example, smaller sub-networks are faster but less accurate than larger ones. At runtime, dynamic DNNs can change the architecture among their sub-networks to comply with new latency targets and dynamically available hardware resources. However, existing works treat dynamic DNNs as an algorithm-only approach, without considering hardware trade-off opportunities. This limits the trade-off range and granularity in latency, power and energy. Furthermore, standalone channel/layer scaling cannot fully utilise heterogeneous computing resources on modern mobile/embedded platforms, and their application is limited to ConvNet \cite{lou2021dynamic}.

\begin{figure}[tbp]
\centerline{\includegraphics[width=\columnwidth]{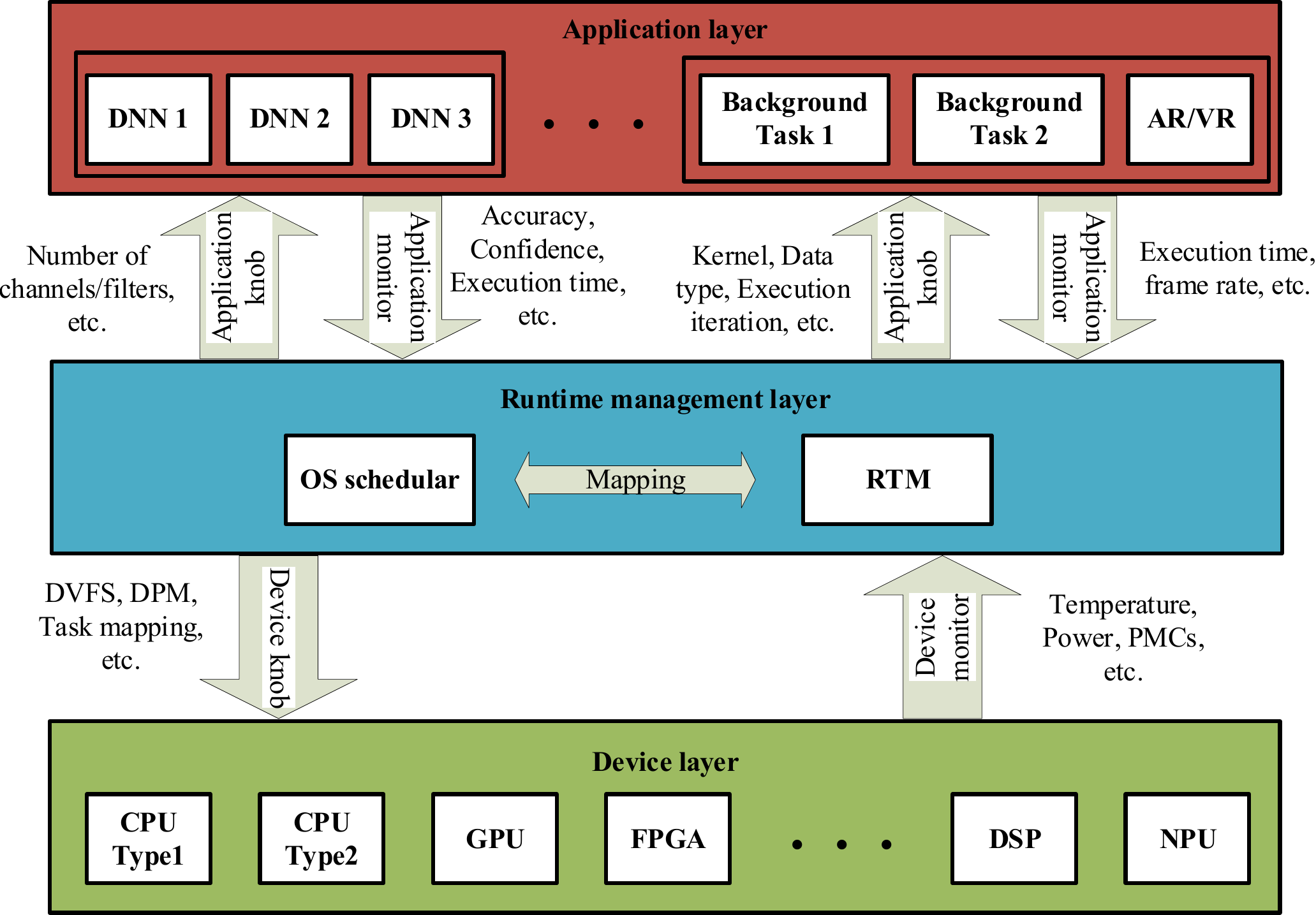}}
\caption{A holistic system design for DNN performance and energy trade-offs. Algorithm knobs (e.g. dynamic DNNs) and hardware knobs (e.g. task mapping, dynamic voltage and frequency scaling (DVFS)) are combined to meet algorithms' dynamically changing performance targets (e.g. accuracy, latency) as well as hardware resources (e.g. available cores, clock frequency level) and constraints (e.g. power, temperature).}
\label{fig1}
\end{figure}

We propose a holistic system design for DNN performance and energy optimisation, combining the trade-off opportunities in both algorithms and hardware. As shown in Fig \ref{fig1}, a system can be viewed as three abstract layers: the device layer contains heterogeneous computing resources; the application layer has multiple concurrent workloads; and the runtime resource management layer monitors the dynamically changing algorithms' performance targets as well as hardware resources and constraints, and tries to meet them by tuning the algorithm and hardware at the same time. Moreover, We illustrate the runtime approach through a dynamic version of `once-for-all network \cite{cai2019once}' (namely Dynamic-OFA), which can scale the ConvNet architecture to fit heterogeneous computing resources efficiently \cite{lou2021dynamic} and has good generalisation for different model architectures such as Transformer \cite{parry2021dynamic}. Compared to the state-of-the-art Dynamic DNNs, our experimental results using ImageNet on a Jetson Xavier NX show that the Dynamic-OFA is up to 3.5x (CPU), 2.4x (GPU) faster for similar ImageNet Top-1 accuracy, or 3.8\% (CPU), 5.1\% (GPU) higher accuracy at similar latency. Furthermore, compared with Linux governor (e.g. performance, schedutil), our runtime approach reduces the energy consumption by 16.5\% at similar latency.

\end{document}